\documentclass[reprint,amsmath,amssymb,aps,pre,floatfix]{revtex4-1}
\usepackage{graphicx}
\usepackage{comment}
\usepackage{dcolumn}
\usepackage{bm}
\usepackage{amsmath}
\usepackage{qcircuit}
\usepackage{color}
\usepackage{xypic}
\usepackage{physics}
\usepackage{hyperref}
\usepackage{float}
\begin{document}

\title{Out of Time Order Correlation of the Hubbard Model with Random Local Disorder}

\author{Chakradhar Rangi}
\affiliation{Department of Physics and Astronomy, Louisiana State University, Baton Rouge, Louisiana 70803, USA}

\author{Juana Moreno}
\affiliation{Department of Physics and Astronomy, Louisiana State University, Baton Rouge, Louisiana 70803, USA}%
\affiliation{Center for Computation and Technology, Louisiana State University, Baton Rouge, LA 70803, USA}

\author{Ka-Ming Tam}
\affiliation{Department of Physics and Astronomy, Louisiana State University, Baton Rouge, Louisiana 70803, USA}%
\affiliation{Center for Computation and Technology, Louisiana State University, Baton Rouge, LA 70803, USA }

\date{\today}

\begin{abstract}

The out-of-time-order correlator (OTOC) serves as a powerful tool for investigating quantum information spreading and chaos in complex systems. We present a method employing non-equilibrium dynamical mean-field theory (DMFT) and coherent potential approximation (CPA) combined with diagrammatic perturbation on the Schwinger-Keldysh contour to calculate the OTOC for correlated fermionic systems subjected to both random disorder and electrons interaction. Our key finding is that random disorder enhances the OTOC decay in the Hubbard model for the metallic phase in the weak coupling limit. However, the current limitation of our perturbative solver restricts the applicability to weak interaction regimes.

\end{abstract}

\maketitle

\section{Introduction}

Understanding how quantum mechanics evolves into the chaotic behavior of classical systems is an active and important research area explored through the field of quantum chaos\cite{stockmann2000quantum,gutzwiller1992quantum,haake1991quantum}. 
An important idea to character chaos in a quantum system lies in the statistical description of energy spectra. Specifically, the distribution of eigenvalue spacing in quantum chaotic systems follows the Gaussian Orthogonal Ensemble (GOE) of random matrices \cite{berry1977level,Periodic2007Heusler,Bohigas_Giannoni_Schmit_1984,McDonald_Kaufman_1979,casati1980connection}. 
However, finding the spectrum is often difficult and an alternative probe of quantum chaos based on the out of time order correlator (OTOC) was proposed.
The capability of the OTOC for probing chaos can be understood from the semi-classical limit. A defining characteristic of chaos is its sensitivity to the initial condition. For a map $x_{i+1}=f(x_{i})$, if the difference in the initial conditions is $\delta x_{0}$, then $\delta x_{i+1} = df / dx |_{x=x_{i}} \delta x_{i}$. After $N$ iterations, the difference can be expressed as $\delta x_{N} = exp(\lambda N) \delta x_{0}$, where $\lambda \approx (1/N) \sum_{i=0}^{N-1} log |df(x)/dx|_{x=x_{i}}$. The $\lambda$ is the Lyapunov exponent, which measures the deviation or sensitivity to the initial condition. For a Hamiltonian dynamical system, the Poisson's bracket (PB) for the generalized coordinates and momentum at different times can be written as  $\{  q(t),p(0)\}_{PB} =  \partial q(t) / \partial q(0) $. If the system is chaotic, one expects that the above PB to grow exponentially, that is $ \{q(t),p(0)\}_{PB} \sim exp(\lambda t)$ \cite{cotler2018out,trunin2021pedagogical,maldacena2016bound}.

Considering the naive semi-classical limit, and substituting the PB with the commutator for quantum mechanics $-(i/\hbar)[q(t),p(0)]$, one could expect that the quantum counterpart of the chaotic system should have an exponential growth in the commutator. The amplitude square of the commutator with respect to the expectation value over an initial state should naively scale as $\langle [q(t),p(0)] [q(t),p(0)]^{\dagger} \rangle \sim exp(2 \lambda t)$. This expectation value of the amplitude square of the commutators first proposed for a semi-classical theory of superconductivity \cite{larkin1969quasiclassical} has recently been used as a measure of the quantum chaos with the corresponding Lyapunov exponent, $\lambda$ \cite{PhysRevLett.115.131603,shenker2014black,maldacena2016remarks}. 

This idea was further generalized for a pair of general operators, denoted as $A$ and $B$, separated by time $t$, $\langle [A(t),B(0)] [A(t),B(0)]^{\dagger} \rangle$ \cite{kitaev2015simple,larkin1969quasiclassical,Hashimoto_2017}. This definition may not always have a direct classical counterpart, for example one can consider $A$ and $B$ as the creation and annihilation operators.
A particular well studied correlator is the OTOC defined as $\langle A(t)^{\dagger}B(0)^{\dagger}A(t)B(0) \rangle$, it has been suggested that this OTOC grows exponentially as $\alpha_{0} - \alpha_{1} exp(\lambda t)$.\cite{Kobrin_etal_2021,SYK_RMP,otoc_dmft_qmc}. 


The OTOC and its variances have been extensively studied for the solvable 
SYK model \cite{SYK_RMP}. However, calculating the OTOC for other interacting quantum systems remains a significant challenge, 
especially when they lack an exact solution. While techniques like direct diagonalization and Krylov subspace 
methods offer numerical solutions, their scalability is often limited to small system sizes \cite{Kobrin_etal_2021}. For systems near 
equilibrium, analytic continuation from Matsubara imaginary-time correlation functions allows OTOC calculation 
using quantum Monte Carlo \cite{otoc_dmft_qmc}, but the infamous "minus sign problem" can still hinder its applicability.

To complement existing methods, we explore a perturbative approach on the double folded Schwinger-Keldysh contour. This approach combines the non-perturbative nature of the dynamical mean field theory (DMFT) with the coherent 
potential approximation (CPA) for disorder averaging \cite{Aoki_2014,Freericks_2019,Freericks_2008,Dohner_2022,dohner2023thermalization,Yan_2023}. 
We will apply this method to the non-equilibrium spin-half 
Anderson-Hubbard model, which lacks integrability and analytical solutions, making traditional numerical 
methods computationally expensive or even infeasible. We note that the OTOC for interacting disorder 
models has been previously studied using the non-linear sigma model \cite{liao2018nonlinear}. Moreover, the strong disorder cases have been studied extensively
by numeric and scaling argument \cite{swingle2017slow,he2017characterizing,chen2016universal}. 

Our study reveals that random disorder enhances the initial OTOC decays. While the data is not sufficient to decide whether this is decaying exponentially due to the limitation of the perturbative method. Thus, we are not able to accurately extract the Lyapunov exponent. 

This paper is organized as follow. In the section II, we briefly review the method for the DMFT 
for systems at non-equilibrium with random disorder. The method for calculating the OTOC is presented in the section III. The results of the OTOC, is presented in the section IV. We conclude and also discuss the possible future works for using DMFT combined with the perturbation method for the study of quantum chaos and quantum information spreading.

\section{Model and DMFT Approximation}
The model we study is the Anderson-Hubbard model with local interaction and local random potential given as 
\begin{eqnarray}\label{eqn:model}
    H = -t \sum_{i,j,\sigma} (c^{\dagger}_{\sigma} c_{\sigma} + h.c.) + U(t)\sum_{i}n_{i,\uparrow}n_{i,\downarrow} \\ \nonumber +\sum_{i} V_{i} (n_{i,\uparrow} + n_{i,\downarrow}) - \mu\sum_{i} n_{i},
\end{eqnarray}
where $V_{i}$ follows the distribution $\rho(V_{i})$. $c^{\dagger}_{i,\sigma}$ and $c_{i,\sigma}$ are creation and annihilation operators for spin one half fermions respectively. $n_{i,\uparrow}$ and $n_{i,\downarrow}$ are  the charge density operators for spin up and spin down  respectively. $n_{i}=n_{i,\uparrow} + n_{i,\downarrow}$ is the total charge density. We only consider bimodal distribution in this study, that is $\rho(V_{i}) = 1/2 [\delta(V_{i}-W)+ \delta(V_{i}+W)]$. $t$ is set to $0.25$ and its serves as the energy scale. We choose $\mu=U/2$ which sets the half-filled case after disorder averaging. Note that we did not try to fix the filling for each individual random realization. The on-site interaction is, in general, a function of time as we consider the non-equilibrium properties of the model. Our strategy is to obtain the non-equilibrium Green's function, $G(t,t^{'})$, for the model under the DMFT approximation and then use perturbation theory to calculate the OTOC. 


The key idea of DMFT lies in mapping the complex problem of interacting electrons on a lattice to an effective single-site problem, the Anderson impurity model. This simplification is valid in the limit of infinite dimensions, it makes the problem significantly more tractable, allowing for numerical and semi-analytical solutions \cite{DMFT_RMP}. A notable success of DMFT is its prediction of the metal-insulator transition in the Hubbard model, a prototypical correlated system \cite{DMFT_RMP}.

Further expanding its reach, DMFT has been generalized to work on the Schwinger-Keldysh contour, enabling the study of non-equilibrium dynamics in correlated systems. Recently, the method has combined with the coherent potential approxiamtion for the study of Anderson-Hubbard model \cite{Dohner_2022,dohner2023thermalization,Yan_2023}. For completeness, we briefly outline the steps of the DMFT here. 

 
The DMFT starts by constructing an effective bath Green's function. This is achieved by spatially averaging the Green's function of the original system, ignoring interactions and disorder. Next, DMFT employs an impurity solver. This solver takes the effective bath Green's function as input, along with information about the local interaction and random local potential, and calculates the Green's function of the interacting impurity problem or the impurity self-energy \cite{DMFT_RMP}.

The impurity Green's function obtained from the solver is used to extract the self-energy via the Schwinger-Dyson equation. This self-energy then serves as an input to calculate the Green's function of the entire lattice using another application of the same equation. Finally, the self-energy and the average Green's function across the lattice are used to update the bath Green's function, closing the loop in the iterative process of DMFT.

To simplify the space averaging procedure involved in obtaining the average Green's function, the model is sometimes placed on a Bethe lattice \cite{bethe1935statistical}. In this study, We consider the model on the Bethe lattice only. 

We summarize the procedure as follow: 
\begin{enumerate}
    \item The spatially averaged bare Green's function for the lattice $G^{lat}_{0}(\vec{k},t,t^{'})$ is used as the bare Green's function for the effective Anderson impurity $G^{imp}_{0}(t,t^{'})$.
    
    \item An impurity solver is employed to solved the effective Anderson impurity problem with bare Green function $G^{imp}_{0}(t,t^{'})$, $U$, and disorder potential $\rho(V)$. The output of the impurity solver can be represented in term of Green's function or self-energy.

    \item Following the coherent potential approximation \cite{Dohner_2022}, the averaged impurity Green's function is given by the averaging of the impurity Green's function with local random disorder potential, $G^{imp}_{ave}(t,t^{'}) = \int dV \rho(V) G^{imp}(V;t,t^{'})$.

    \item With the averaged impurity Green's function or self-energy $\Sigma_{ave}^{imp}(t,t^{'})$, the Schwinger-Dyson equation is used to generate the lattice Green's function $G^{lat}(\vec{k},t,t^{'}) = ((G^{lat}_{0}(\vec{k},t,t^{'}))^{-1} - \Sigma_{ave}^{imp}(t,t^{'}))^{-1}$
    
    \item The lattice Green's function is then spatially averaged to form the bare Green's function via the Schwinger-Dyson equation for the effective Anderson impurity and this completes the self-consistency loop. 
\end{enumerate}

We use the second order perturbation to calculate the self-energy of the effective Anderson impurity problem in step 2 \cite{Aoki_2014}. The first order contributions include the Hartree and the Fock terms. The Fock term is zero for the Hubbard model and the Hatree term is also zero due to the condition $\mu=U/2$. After, the self-consistency is attained. The bare Green's function for the effective Anderson impurity $G_{0}^{imp}(t,t')$ is then used to compute the OTOC using the diagrammatic perturbation theory.

\section{Perturbation theory for calculating OTOC}

\begin{figure}[htbp]
            \centering
            \begin{minipage}{\columnwidth}
                \centering
                \includegraphics[width=70mm,scale=0.5]{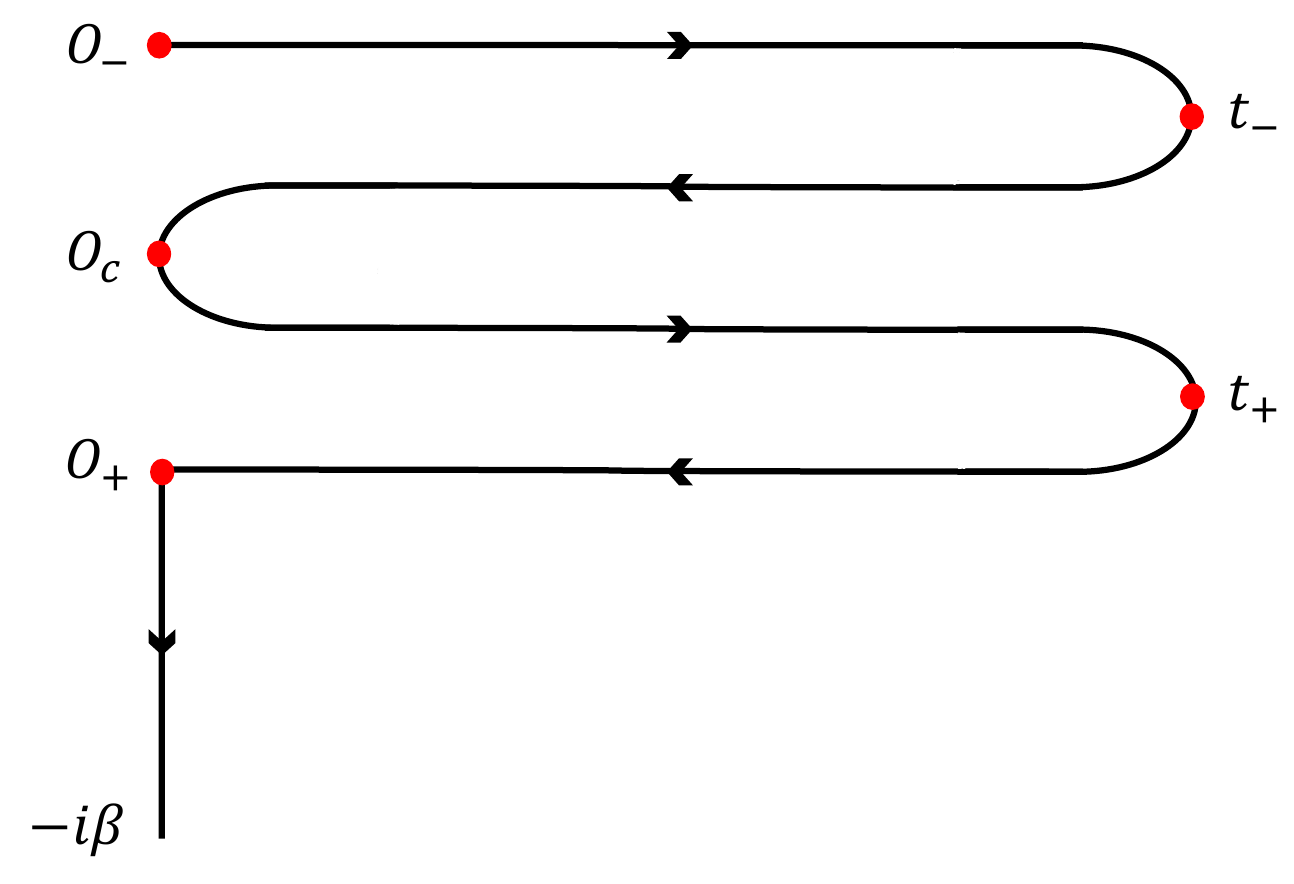} 
            \end{minipage}
            \caption{Double folded Schwinger-Keldysh contour used in the DMFT. The first fold extends from $0_{-}$ to $0_{c}$ (the time between these two zeros is denoted as $t_{-}$), the second fold extends from $0_{c}$ to $0_{+}$ (the time between these two zeros is denoted as $t_{+}$), and from $0_{+}$ to $-i\beta $ is the thermal branch.}            \label{fig:4foldKeldysh}
\end{figure}

For usual physical quantities, such as susceptibility, the response function involves time-ordered products of operators. Standard diagrammatic perturbation theory relies on decomposing these time-ordered operators into products of Green's functions using Wick's theorem. However, it is not directly applicable for product of operators without the time-ordered structure. A straightforward solution is to generalize the Schwinger-Keldysh contour into multiple folds, allowing operators to be ordered along the contour time, see Fig. \ref{fig:4foldKeldysh}.

The multi-fold Schwinger-Keldysh contour has received significant recent attention and has been applied with DMFT to calculate the OTOC for the Falicov-Kimball (FK) model \cite{Tsuji_2017}. 
Notably, the FK model allows for an exact solution within the DMFT framework \cite{Freericks_Zlatic_2003}, enabling the precise calculation of the OTOC \cite{Tsuji_2017}. In contrast, our current focus is on utilizing perturbation theory to 
calculate the OTOC for quantum interacting models that lack such an exact solution.

\begin{figure}[htbp]
            \centering
            \begin{minipage}{\columnwidth}
                \centering
                \includegraphics[width=75mm]{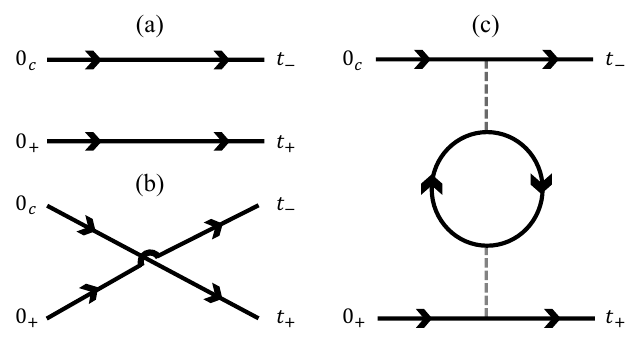} 
            \end{minipage}
            \caption{The zeroth order and the second order diagrams for the calculation of OTOC. The dashed line is the Hubbard vertex. Since the external legs have the same spin, combined with the fact that the Hubbard vertex only coupled spin up with spin down electrons, there is only one diagram in the second order. }
            \label{fig:Diagrams}
\end{figure}

We define the real time OTOC and write it in term of the path integral over the double folded Schwinger-Kelydsh contour as follow:
\begin{align}\label{eqn:OTOC}
    F(t) &= \expval{c^\dagger(t_-)c(0_c)c^\dagger(t_+)c(0_+)} \\
    &= \frac{1}{\mathcal{Z}}\int \mathcal{D}[\psi,\Bar{\psi}] \Bar{\psi}(t_-)\psi(0_c)\Bar{\psi}(t_+)\psi(0_+) e^{i\mathcal{S[\psi,\Bar{\psi}]}},
\end{align}
where $\psi$ and $\Bar{\psi}$ are Grassmann fields, $\mathcal{S}$ is the action, $\mathcal{D}[\psi,\Bar{\psi}]$ is the measure of the path integral, the partition function $\mathcal{Z}$ is given by $\mathcal{Z} = \int \mathcal{D}[\psi,\Bar{\psi}]e^{i\mathcal{S[\psi,\Bar{\psi}]}}$, and the lattice action given by our model in Eq.\eqref{eqn:model}.  Since the operators are now contour time ordered, we can use the Wick's theorem to perturbatively compute the various diagrams contributing to $F(t)$. All the non-vanishing diagrams up to second order in $U$ are given in Fig. \ref{fig:Diagrams}. 
We note that only spin up component is calculated, as the DMFT we employed here does not break spin rotational symmetry and has no magnetic ordering. 

At the zeroth order, two Wick's contractions are possible as illustrated in Fig. \hyperref[fig:Diagrams]{2(a)} and \hyperref[fig:Diagrams]{2(b)}. Moving to the first order, the diagrams are proportional to the Hartree term. However, the Hartree term vanishes  due to the averaged half-filling condition ($\mu=U/2$). At the second order, only connected diagrams, i.e., one-particle irreducible diagrams, are taken into account, resulting in the diagram shown in Fig. \hyperref[fig:Diagrams]{2(c)}.

For practical calculation the Schwinger-Keldysh contour is discretized into finite number of time steps along the entire contour. Under this approximation, the two time non-equilibrium Green's function can be represented in term of a finite size $N \times N$ matrix, where $N$ is the total number of time steps on the contour. There are four temporal branches and one thermal branch, that is $N=4N_{t}+N_{\tau}$. In this paper, we use $N_t = 500$ steps on the real-time branch and $N_\tau=100$ steps on the thermal branch. The bare Green's function can be found by finding the inverse of the $(i\partial_t + \mu) \delta(t,t^{'}) = G_{0}(t,t^{'})$. Schwinger-Dyson equation can also be defined along the contour. We simply generalized the numerical implementation in Ref. \onlinecite{Freericks_2008} for the double folded contour. We note that implementation which use the Runge-Kutta solver is also a popular method for solving the bare Green's function on the Schwinger-Kelydsh contour \cite{Eckstein_Werner_2011,stan2009time,Tsuji_2017}. Once the DMFT cycle is converged, the Green's function on the double folded Schwinger-Keldysh contour are then extracted for calculating the diagrams for the OTOC up to the second order in $U$.

\section{Results}
\begin{figure}[htbp]
            \centering
            \begin{minipage}{\columnwidth}
                \centering
                \includegraphics[width=1.0\columnwidth]{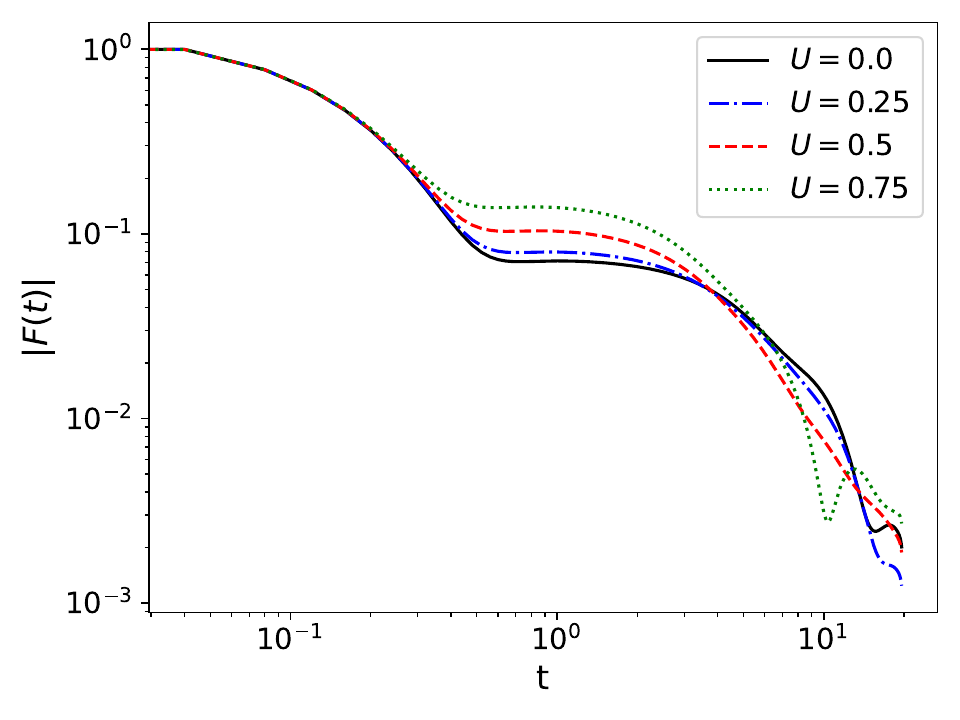} 
            \end{minipage}
            \caption{The modulus of OTOC $|F(t)|$ (normalized at $t=0$) [Eq. \eqref{eqn:OTOC}], on a log-log scale, of the Anderson-Hubbard model in the clean system limit $W=0$ for various values of the interaction strength $U$. The values of $U$ correspond to the  metallic phase of the Hubbard model. Only the spin up component is calculated as the DMFT does not break spin rotational symmetry}
            \label{fig:LogLogPlotOOTCVaryingU}
\end{figure}

We compute and discuss the OTOC obtained from the method described above. The system is initially at the thermal equilibrium with temperature $T=1/\beta$ with $\beta=100$ and maximum simulation time $t_{\text{max}}$ is set to $t_{\text{max}}=20$. The interaction $U$ and the disorder strength $W$ are kept constant in both the thermal and real branches of the contour. We plot the OTOC for different values of interaction strengths as shown in Fig. \ref{fig:LogLogPlotOOTCVaryingU}. 
Across all investigated interaction strengths, the systems are always in the metallic phase. The OTOC does not reveal any discernible trend with increasing interaction within the explored time window.


\begin{figure}[htbp]
            \centering
            \includegraphics[width=1.0\columnwidth]{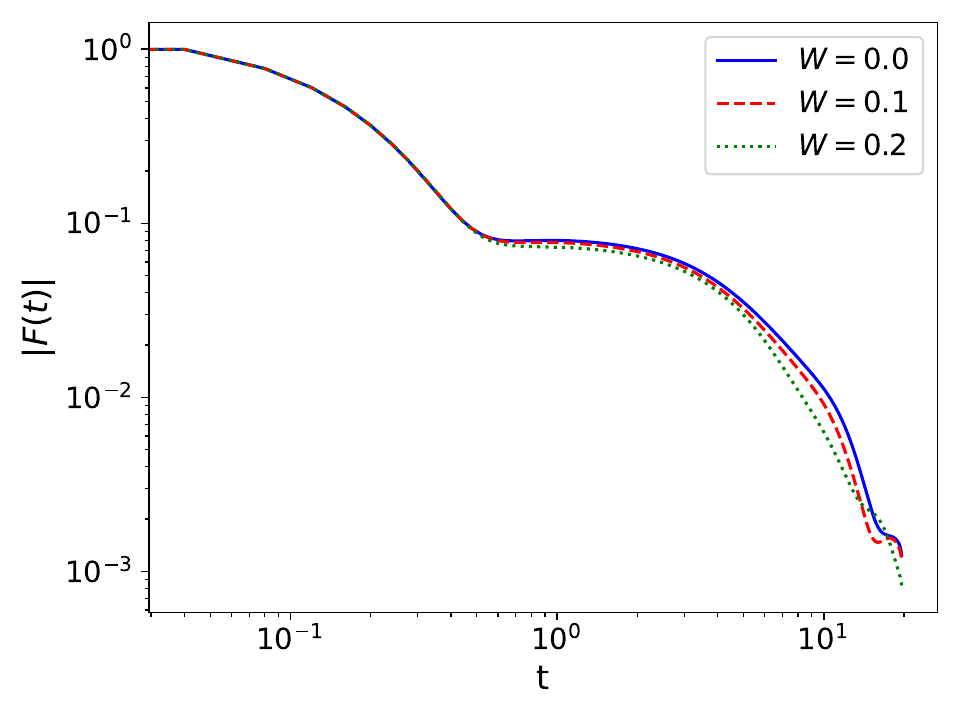}
            \includegraphics[width=1.0\columnwidth]{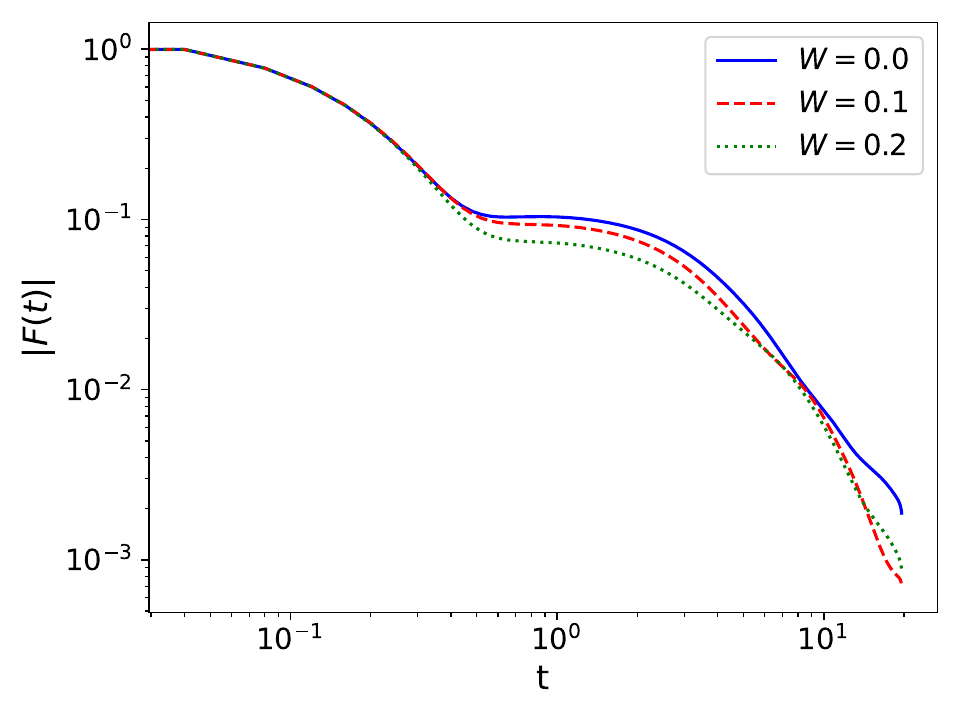} 
            \includegraphics[width=1.0\columnwidth]{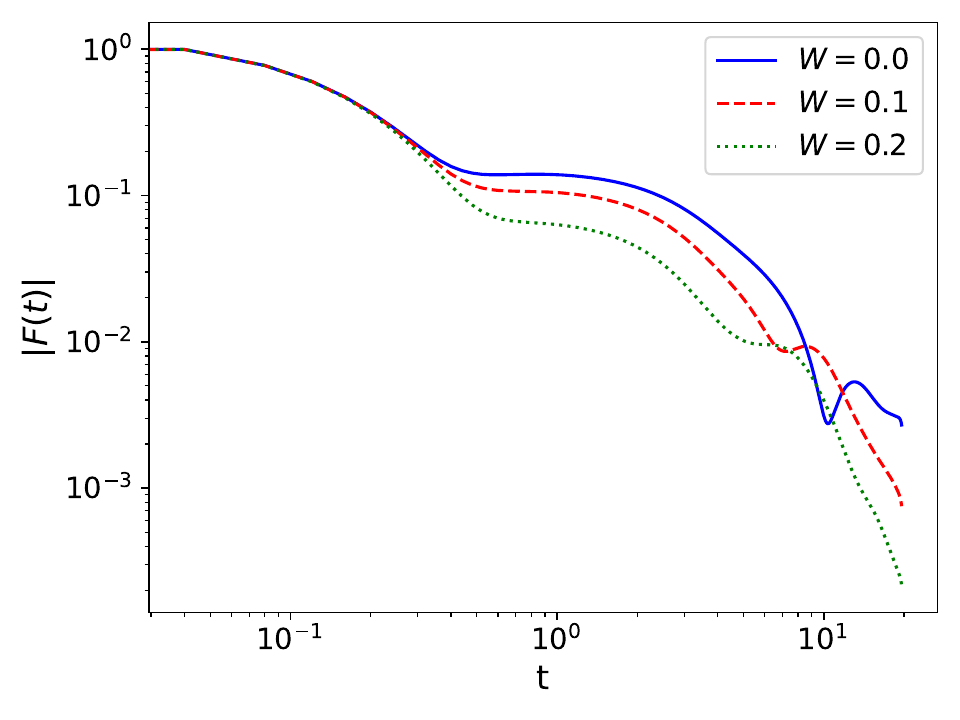}
            \caption{The modulus of the OTOC $|F(t)|$(normalized at $t=0$) on a log-log scale for different values of the interaction strength $U$ and disorder strength $W$. (a) $U=0.25$ (b) $U=0.5$   (c) $U=0.75$ . The solid lines in (a), (b) and (c) denote clean system limit $W=0$, the dashed line $W=0.1$, and the dotted line $W=0.2$. The DMFT we employed here does not break spin rotational symmetry, thus we only calculate the spin up component for the OTOC.}
            \label{fig:LogLogPlotOOTCVaryingW}
\end{figure}

The OTOC for the Hubbard model with random disorder are plotted in the Fig.\ref{fig:LogLogPlotOOTCVaryingW}. Three different values of the interaction strength, $U=0.25, 0.5$ and $0.75$, in each case two values of disorder strength $0.1$ and $0.2$ together with the disorder free case ($W=0$) are shown.

The main effect from the random disorder is to drive the OTOC to decay faster. Ideally, we could fit the data to extract the Lyapunov exponent and quantify this effect more precisely. However, current limitations in the maximum time hinder accurate fitting. Nonetheless, these results suggest that weak disorder may accelerate thermalization and information scrambling in correlated fermionic systems based on the accelerated decays of the OTOC from the disorder. 

\section{Conclusion}
While exactly solvable models like the SYK model offer valuable insights, particularly regarding their 
connection to holographic duality, alternative methods are necessary to study the vast majority of systems 
that lack such solutions. The Hubbard model, despite its importance in condensed matter physics, 
remains particularly challenging for calculating temporal dynamics such as OTOC. Although DMFT proves 
accurate and computational tractable for systems at equilibrium. The non-equilibrium DMFT, in particular for calculating two particles and higher order Green's functions, is generally rather difficult \cite{Aoki_2014}. 

This work proposes a prescription for calculating OTOC by combining perturbation theory with the DMFT 
framework, specifically tailored for such challenging models. Our approach extends the possibility of 
investigating the OTOC for a wider range of models, enabling a more comprehensive understanding of 
non-equilibrium dynamics in systems with interplay between electrons correlation and random disorder.

The main conclusion from our results for the Hubbard model at weak interaction is that disorder tends 
to accelerate the decay of the OTOC as a function of time. Unfortunately, at present we are not able to capture the competition between Mott insulator and Anderson insulator at strong coupling and strong disorder regimes. 

Even at weak coupling, our current data is insufficient to accurately determine the decay exponent or definitively confirm whether the decay is exponential.
There are two directions for improving the method. First, the impurity solver for the DMFT iteration can be improved by going beyond second order perturbation. Second, partial summation of diagrams for calculating the OTOC, particularly ring diagrams to all orders, could be explored.

As discussed above, the current approach has limitations due to the perturbative nature of the impurity solver and the lack of spatial correlations within DMFT. Despite these limitations, the flexibility of the perturbative approach allows us to investigate various intriguing systems. This includes studying the effects of quenching interaction and even quenching disorder, along with temperature dependence of the OTOC.

\section{Acknowledgment}
 We are grateful to Herbert Fotso and Hanna Terletska for useful discussions on the related projects.
 This work used high-performance computational resources provided by the Louisiana Optical Network Initiative (http://www.loni.org) and HPC@LSU computing. JM is supported by the US Department of Energy, Office of Science, Office of Basic Energy Sciences, under Award Number DE-SC0017861.


\bibliography{references.bib}

\begin{thebibliography}{36}%
\makeatletter
\providecommand \@ifxundefined [1]{%
 \@ifx{#1\undefined}
}%
\providecommand \@ifnum [1]{%
 \ifnum #1\expandafter \@firstoftwo
 \else \expandafter \@secondoftwo
 \fi
}%
\providecommand \@ifx [1]{%
 \ifx #1\expandafter \@firstoftwo
 \else \expandafter \@secondoftwo
 \fi
}%
\providecommand \natexlab [1]{#1}%
\providecommand \enquote  [1]{``#1''}%
\providecommand \bibnamefont  [1]{#1}%
\providecommand \bibfnamefont [1]{#1}%
\providecommand \citenamefont [1]{#1}%
\providecommand \href@noop [0]{\@secondoftwo}%
\providecommand \href [0]{\begingroup \@sanitize@url \@href}%
\providecommand \@href[1]{\@@startlink{#1}\@@href}%
\providecommand \@@href[1]{\endgroup#1\@@endlink}%
\providecommand \@sanitize@url [0]{\catcode `\\12\catcode `\$12\catcode `\&12\catcode `\#12\catcode `\^12\catcode `\_12\catcode `\%12\relax}%
\providecommand \@@startlink[1]{}%
\providecommand \@@endlink[0]{}%
\providecommand \url  [0]{\begingroup\@sanitize@url \@url }%
\providecommand \@url [1]{\endgroup\@href {#1}{\urlprefix }}%
\providecommand \urlprefix  [0]{URL }%
\providecommand \Eprint [0]{\href }%
\providecommand \doibase [0]{http://dx.doi.org/}%
\providecommand \selectlanguage [0]{\@gobble}%
\providecommand \bibinfo  [0]{\@secondoftwo}%
\providecommand \bibfield  [0]{\@secondoftwo}%
\providecommand \translation [1]{[#1]}%
\providecommand \BibitemOpen [0]{}%
\providecommand \bibitemStop [0]{}%
\providecommand \bibitemNoStop [0]{.\EOS\space}%
\providecommand \EOS [0]{\spacefactor3000\relax}%
\providecommand \BibitemShut  [1]{\csname bibitem#1\endcsname}%
\let\auto@bib@innerbib\@empty
\bibitem [{\citenamefont {St{\"o}ckmann}(2000)}]{stockmann2000quantum}%
  \BibitemOpen
  \bibfield  {author} {\bibinfo {author} {\bibfnamefont {H.-J.}\ \bibnamefont {St{\"o}ckmann}},\ }\href@noop {} {\enquote {\bibinfo {title} {Quantum chaos: an introduction},}\ } (\bibinfo {year} {2000})\BibitemShut {NoStop}%
\bibitem [{\citenamefont {Gutzwiller}(1992)}]{gutzwiller1992quantum}%
  \BibitemOpen
  \bibfield  {author} {\bibinfo {author} {\bibfnamefont {M.~C.}\ \bibnamefont {Gutzwiller}},\ }\href@noop {} {\bibfield  {journal} {\bibinfo  {journal} {Scientific American}\ }\textbf {\bibinfo {volume} {266}},\ \bibinfo {pages} {78} (\bibinfo {year} {1992})}\BibitemShut {NoStop}%
\bibitem [{\citenamefont {Haake}(1991)}]{haake1991quantum}%
  \BibitemOpen
  \bibfield  {author} {\bibinfo {author} {\bibfnamefont {F.}~\bibnamefont {Haake}},\ }\href@noop {} {\emph {\bibinfo {title} {Quantum signatures of chaos}}}\ (\bibinfo  {publisher} {Springer},\ \bibinfo {year} {1991})\BibitemShut {NoStop}%
\bibitem [{\citenamefont {Berry}\ and\ \citenamefont {Tabor}(1977)}]{berry1977level}%
  \BibitemOpen
  \bibfield  {author} {\bibinfo {author} {\bibfnamefont {M.~V.}\ \bibnamefont {Berry}}\ and\ \bibinfo {author} {\bibfnamefont {M.}~\bibnamefont {Tabor}},\ }\href@noop {} {\bibfield  {journal} {\bibinfo  {journal} {Proc. R. Soc. A}\ }\textbf {\bibinfo {volume} {356}},\ \bibinfo {pages} {375} (\bibinfo {year} {1977})}\BibitemShut {NoStop}%
\bibitem [{\citenamefont {Heusler}\ \emph {et~al.}(2007)\citenamefont {Heusler}, \citenamefont {M\"uller}, \citenamefont {Altland}, \citenamefont {Braun},\ and\ \citenamefont {Haake}}]{Periodic2007Heusler}%
  \BibitemOpen
  \bibfield  {author} {\bibinfo {author} {\bibfnamefont {S.}~\bibnamefont {Heusler}}, \bibinfo {author} {\bibfnamefont {S.}~\bibnamefont {M\"uller}}, \bibinfo {author} {\bibfnamefont {A.}~\bibnamefont {Altland}}, \bibinfo {author} {\bibfnamefont {P.}~\bibnamefont {Braun}}, \ and\ \bibinfo {author} {\bibfnamefont {F.}~\bibnamefont {Haake}},\ }\href {\doibase 10.1103/PhysRevLett.98.044103} {\bibfield  {journal} {\bibinfo  {journal} {Phys. Rev. Lett.}\ }\textbf {\bibinfo {volume} {98}},\ \bibinfo {pages} {044103} (\bibinfo {year} {2007})}\BibitemShut {NoStop}%
\bibitem [{\citenamefont {Bohigas}\ \emph {et~al.}(1984)\citenamefont {Bohigas}, \citenamefont {Giannoni},\ and\ \citenamefont {Schmit}}]{Bohigas_Giannoni_Schmit_1984}%
  \BibitemOpen
  \bibfield  {author} {\bibinfo {author} {\bibfnamefont {O.}~\bibnamefont {Bohigas}}, \bibinfo {author} {\bibfnamefont {M.~J.}\ \bibnamefont {Giannoni}}, \ and\ \bibinfo {author} {\bibfnamefont {C.}~\bibnamefont {Schmit}},\ }\href {\doibase 10.1103/PhysRevLett.52.1} {\bibfield  {journal} {\bibinfo  {journal} {Phys. Rev. Lett.}\ }\textbf {\bibinfo {volume} {52}},\ \bibinfo {pages} {1} (\bibinfo {year} {1984})}\BibitemShut {NoStop}%
\bibitem [{\citenamefont {McDonald}\ and\ \citenamefont {Kaufman}(1979)}]{McDonald_Kaufman_1979}%
  \BibitemOpen
  \bibfield  {author} {\bibinfo {author} {\bibfnamefont {S.~W.}\ \bibnamefont {McDonald}}\ and\ \bibinfo {author} {\bibfnamefont {A.~N.}\ \bibnamefont {Kaufman}},\ }\href {\doibase 10.1103/PhysRevLett.42.1189} {\bibfield  {journal} {\bibinfo  {journal} {Phys. Rev. Lett.}\ }\textbf {\bibinfo {volume} {42}},\ \bibinfo {pages} {1189} (\bibinfo {year} {1979})}\BibitemShut {NoStop}%
\bibitem [{\citenamefont {Casati}\ \emph {et~al.}(1980)\citenamefont {Casati}, \citenamefont {Valz-Gris},\ and\ \citenamefont {Guarnieri}}]{casati1980connection}%
  \BibitemOpen
  \bibfield  {author} {\bibinfo {author} {\bibfnamefont {G.}~\bibnamefont {Casati}}, \bibinfo {author} {\bibfnamefont {F.}~\bibnamefont {Valz-Gris}}, \ and\ \bibinfo {author} {\bibfnamefont {I.}~\bibnamefont {Guarnieri}},\ }\href@noop {} {\bibfield  {journal} {\bibinfo  {journal} {Lett. Nuovo Cim. (1971-1985)}\ }\textbf {\bibinfo {volume} {28}},\ \bibinfo {pages} {279} (\bibinfo {year} {1980})}\BibitemShut {NoStop}%
\bibitem [{\citenamefont {Cotler}\ \emph {et~al.}(2018)\citenamefont {Cotler}, \citenamefont {Ding},\ and\ \citenamefont {Penington}}]{cotler2018out}%
  \BibitemOpen
  \bibfield  {author} {\bibinfo {author} {\bibfnamefont {J.~S.}\ \bibnamefont {Cotler}}, \bibinfo {author} {\bibfnamefont {D.}~\bibnamefont {Ding}}, \ and\ \bibinfo {author} {\bibfnamefont {G.~R.}\ \bibnamefont {Penington}},\ }\href@noop {} {\bibfield  {journal} {\bibinfo  {journal} {Ann. Phys.}\ }\textbf {\bibinfo {volume} {396}},\ \bibinfo {pages} {318} (\bibinfo {year} {2018})}\BibitemShut {NoStop}%
\bibitem [{\citenamefont {Trunin}(2021)}]{trunin2021pedagogical}%
  \BibitemOpen
  \bibfield  {author} {\bibinfo {author} {\bibfnamefont {D.~A.}\ \bibnamefont {Trunin}},\ }\href@noop {} {\bibfield  {journal} {\bibinfo  {journal} {Physics-Uspekhi}\ }\textbf {\bibinfo {volume} {64}},\ \bibinfo {pages} {219} (\bibinfo {year} {2021})}\BibitemShut {NoStop}%
\bibitem [{\citenamefont {Maldacena}\ \emph {et~al.}(2016)\citenamefont {Maldacena}, \citenamefont {Shenker},\ and\ \citenamefont {Stanford}}]{maldacena2016bound}%
  \BibitemOpen
  \bibfield  {author} {\bibinfo {author} {\bibfnamefont {J.}~\bibnamefont {Maldacena}}, \bibinfo {author} {\bibfnamefont {S.~H.}\ \bibnamefont {Shenker}}, \ and\ \bibinfo {author} {\bibfnamefont {D.}~\bibnamefont {Stanford}},\ }\href@noop {} {\bibfield  {journal} {\bibinfo  {journal} {J. High Energy Phys.}\ }\textbf {\bibinfo {volume} {2016}},\ \bibinfo {pages} {1} (\bibinfo {year} {2016})}\BibitemShut {NoStop}%
\bibitem [{\citenamefont {Larkin}\ and\ \citenamefont {Ovchinnikov}(1969)}]{larkin1969quasiclassical}%
  \BibitemOpen
  \bibfield  {author} {\bibinfo {author} {\bibfnamefont {A.~I.}\ \bibnamefont {Larkin}}\ and\ \bibinfo {author} {\bibfnamefont {Y.~N.}\ \bibnamefont {Ovchinnikov}},\ }\href@noop {} {\bibfield  {journal} {\bibinfo  {journal} {Sov Phys JETP}\ }\textbf {\bibinfo {volume} {28}},\ \bibinfo {pages} {1200} (\bibinfo {year} {1969})}\BibitemShut {NoStop}%
\bibitem [{\citenamefont {Roberts}\ and\ \citenamefont {Stanford}(2015)}]{PhysRevLett.115.131603}%
  \BibitemOpen
  \bibfield  {author} {\bibinfo {author} {\bibfnamefont {D.~A.}\ \bibnamefont {Roberts}}\ and\ \bibinfo {author} {\bibfnamefont {D.}~\bibnamefont {Stanford}},\ }\href {\doibase 10.1103/PhysRevLett.115.131603} {\bibfield  {journal} {\bibinfo  {journal} {Phys. Rev. Lett.}\ }\textbf {\bibinfo {volume} {115}},\ \bibinfo {pages} {131603} (\bibinfo {year} {2015})}\BibitemShut {NoStop}%
\bibitem [{\citenamefont {Shenker}\ and\ \citenamefont {Stanford}(2014)}]{shenker2014black}%
  \BibitemOpen
  \bibfield  {author} {\bibinfo {author} {\bibfnamefont {S.~H.}\ \bibnamefont {Shenker}}\ and\ \bibinfo {author} {\bibfnamefont {D.}~\bibnamefont {Stanford}},\ }\href@noop {} {\bibfield  {journal} {\bibinfo  {journal} {J. High Energy Phys.}\ }\textbf {\bibinfo {volume} {2014}},\ \bibinfo {pages} {1} (\bibinfo {year} {2014})}\BibitemShut {NoStop}%
\bibitem [{\citenamefont {Maldacena}\ and\ \citenamefont {Stanford}(2016)}]{maldacena2016remarks}%
  \BibitemOpen
  \bibfield  {author} {\bibinfo {author} {\bibfnamefont {J.}~\bibnamefont {Maldacena}}\ and\ \bibinfo {author} {\bibfnamefont {D.}~\bibnamefont {Stanford}},\ }\href@noop {} {\bibfield  {journal} {\bibinfo  {journal} {Phys. Rev. D}\ }\textbf {\bibinfo {volume} {94}},\ \bibinfo {pages} {106002} (\bibinfo {year} {2016})}\BibitemShut {NoStop}%
\bibitem [{\citenamefont {Kitaev}()}]{kitaev2015simple}%
  \BibitemOpen
  \bibfield  {author} {\bibinfo {author} {\bibfnamefont {A.}~\bibnamefont {Kitaev}},\ }\href@noop {} {\bibinfo  {journal} {Entanglement in Strongly-Correlated Quantum Matter}\ }\BibitemShut {NoStop}%
\bibitem [{\citenamefont {Hashimoto}\ \emph {et~al.}(2017)\citenamefont {Hashimoto}, \citenamefont {Murata},\ and\ \citenamefont {Yoshii}}]{Hashimoto_2017}%
  \BibitemOpen
\bibfield  {journal} {  }\bibfield  {author} {\bibinfo {author} {\bibfnamefont {K.}~\bibnamefont {Hashimoto}}, \bibinfo {author} {\bibfnamefont {K.}~\bibnamefont {Murata}}, \ and\ \bibinfo {author} {\bibfnamefont {R.}~\bibnamefont {Yoshii}},\ }\href {\doibase 10.1007/jhep10(2017)138} {\bibfield  {journal} {\bibinfo  {journal} {Journal of High Energy Physics}\ }\textbf {\bibinfo {volume} {2017}} (\bibinfo {year} {2017}),\ 10.1007/jhep10(2017)138}\BibitemShut {NoStop}%
\bibitem [{\citenamefont {Kobrin}\ \emph {et~al.}(2021)\citenamefont {Kobrin}, \citenamefont {Yang}, \citenamefont {Kahanamoku-Meyer}, \citenamefont {Olund}, \citenamefont {Moore}, \citenamefont {Stanford},\ and\ \citenamefont {Yao}}]{Kobrin_etal_2021}%
  \BibitemOpen
  \bibfield  {author} {\bibinfo {author} {\bibfnamefont {B.}~\bibnamefont {Kobrin}}, \bibinfo {author} {\bibfnamefont {Z.}~\bibnamefont {Yang}}, \bibinfo {author} {\bibfnamefont {G.~D.}\ \bibnamefont {Kahanamoku-Meyer}}, \bibinfo {author} {\bibfnamefont {C.~T.}\ \bibnamefont {Olund}}, \bibinfo {author} {\bibfnamefont {J.~E.}\ \bibnamefont {Moore}}, \bibinfo {author} {\bibfnamefont {D.}~\bibnamefont {Stanford}}, \ and\ \bibinfo {author} {\bibfnamefont {N.~Y.}\ \bibnamefont {Yao}},\ }\href {\doibase 10.1103/PhysRevLett.126.030602} {\bibfield  {journal} {\bibinfo  {journal} {Phys. Rev. Lett.}\ }\textbf {\bibinfo {volume} {126}},\ \bibinfo {pages} {030602} (\bibinfo {year} {2021})}\BibitemShut {NoStop}%
\bibitem [{\citenamefont {Chowdhury}\ \emph {et~al.}(2022)\citenamefont {Chowdhury}, \citenamefont {Georges}, \citenamefont {Parcollet},\ and\ \citenamefont {Sachdev}}]{SYK_RMP}%
  \BibitemOpen
  \bibfield  {author} {\bibinfo {author} {\bibfnamefont {D.}~\bibnamefont {Chowdhury}}, \bibinfo {author} {\bibfnamefont {A.}~\bibnamefont {Georges}}, \bibinfo {author} {\bibfnamefont {O.}~\bibnamefont {Parcollet}}, \ and\ \bibinfo {author} {\bibfnamefont {S.}~\bibnamefont {Sachdev}},\ }\href {\doibase 10.1103/RevModPhys.94.035004} {\bibfield  {journal} {\bibinfo  {journal} {Rev. Mod. Phys.}\ }\textbf {\bibinfo {volume} {94}},\ \bibinfo {pages} {035004} (\bibinfo {year} {2022})}\BibitemShut {NoStop}%
\bibitem [{\citenamefont {Tsuji}\ and\ \citenamefont {Werner}(2019)}]{otoc_dmft_qmc}%
  \BibitemOpen
  \bibfield  {author} {\bibinfo {author} {\bibfnamefont {N.}~\bibnamefont {Tsuji}}\ and\ \bibinfo {author} {\bibfnamefont {P.}~\bibnamefont {Werner}},\ }\href {\doibase 10.1103/PhysRevB.99.115132} {\bibfield  {journal} {\bibinfo  {journal} {Phys. Rev. B}\ }\textbf {\bibinfo {volume} {99}},\ \bibinfo {pages} {115132} (\bibinfo {year} {2019})}\BibitemShut {NoStop}%
\bibitem [{\citenamefont {Aoki}\ \emph {et~al.}(2014)\citenamefont {Aoki}, \citenamefont {Tsuji}, \citenamefont {Eckstein}, \citenamefont {Kollar}, \citenamefont {Oka},\ and\ \citenamefont {Werner}}]{Aoki_2014}%
  \BibitemOpen
  \bibfield  {author} {\bibinfo {author} {\bibfnamefont {H.}~\bibnamefont {Aoki}}, \bibinfo {author} {\bibfnamefont {N.}~\bibnamefont {Tsuji}}, \bibinfo {author} {\bibfnamefont {M.}~\bibnamefont {Eckstein}}, \bibinfo {author} {\bibfnamefont {M.}~\bibnamefont {Kollar}}, \bibinfo {author} {\bibfnamefont {T.}~\bibnamefont {Oka}}, \ and\ \bibinfo {author} {\bibfnamefont {P.}~\bibnamefont {Werner}},\ }\href {\doibase 10.1103/revmodphys.86.779} {\bibfield  {journal} {\bibinfo  {journal} {Rev. Mod. Phys.}\ }\textbf {\bibinfo {volume} {86}},\ \bibinfo {pages} {779} (\bibinfo {year} {2014})}\BibitemShut {NoStop}%
\bibitem [{\citenamefont {Freericks}(2019)}]{Freericks_2019}%
  \BibitemOpen
  \bibfield  {author} {\bibinfo {author} {\bibfnamefont {J.~K.}\ \bibnamefont {Freericks}},\ }\href@noop {} {\bibfield  {journal} {\bibinfo  {journal} {arXiv preprint arXiv:1907.11302}\ } (\bibinfo {year} {2019})}\BibitemShut {NoStop}%
\bibitem [{\citenamefont {Freericks}(2008)}]{Freericks_2008}%
  \BibitemOpen
  \bibfield  {author} {\bibinfo {author} {\bibfnamefont {J.~K.}\ \bibnamefont {Freericks}},\ }\href {\doibase 10.1103/physrevb.77.075109} {\bibfield  {journal} {\bibinfo  {journal} {Phys. Rev. B}\ }\textbf {\bibinfo {volume} {77}} (\bibinfo {year} {2008}),\ 10.1103/physrevb.77.075109}\BibitemShut {NoStop}%
\bibitem [{\citenamefont {Dohner}\ \emph {et~al.}(2022)\citenamefont {Dohner}, \citenamefont {Terletska}, \citenamefont {Tam}, \citenamefont {Moreno},\ and\ \citenamefont {Fotso}}]{Dohner_2022}%
  \BibitemOpen
  \bibfield  {author} {\bibinfo {author} {\bibfnamefont {E.}~\bibnamefont {Dohner}}, \bibinfo {author} {\bibfnamefont {H.}~\bibnamefont {Terletska}}, \bibinfo {author} {\bibfnamefont {K.-M.}\ \bibnamefont {Tam}}, \bibinfo {author} {\bibfnamefont {J.}~\bibnamefont {Moreno}}, \ and\ \bibinfo {author} {\bibfnamefont {H.~F.}\ \bibnamefont {Fotso}},\ }\href {\doibase 10.1103/physrevb.106.195156} {\bibfield  {journal} {\bibinfo  {journal} {Phys. Rev. B}\ }\textbf {\bibinfo {volume} {106}} (\bibinfo {year} {2022}),\ 10.1103/physrevb.106.195156}\BibitemShut {NoStop}%
\bibitem [{\citenamefont {Dohner}\ \emph {et~al.}(2023)\citenamefont {Dohner}, \citenamefont {Terletska},\ and\ \citenamefont {Fotso}}]{dohner2023thermalization}%
  \BibitemOpen
  \bibfield  {author} {\bibinfo {author} {\bibfnamefont {E.}~\bibnamefont {Dohner}}, \bibinfo {author} {\bibfnamefont {H.}~\bibnamefont {Terletska}}, \ and\ \bibinfo {author} {\bibfnamefont {H.~F.}\ \bibnamefont {Fotso}},\ }\href {\doibase 10.1103/PhysRevB.108.144202} {\bibfield  {journal} {\bibinfo  {journal} {Phys. Rev. B}\ }\textbf {\bibinfo {volume} {108}},\ \bibinfo {pages} {144202} (\bibinfo {year} {2023})}\BibitemShut {NoStop}%
\bibitem [{\citenamefont {Yan}\ and\ \citenamefont {Werner}(2023)}]{Yan_2023}%
  \BibitemOpen
  \bibfield  {author} {\bibinfo {author} {\bibfnamefont {J.}~\bibnamefont {Yan}}\ and\ \bibinfo {author} {\bibfnamefont {P.}~\bibnamefont {Werner}},\ }\href {\doibase 10.1103/physrevb.108.125143} {\bibfield  {journal} {\bibinfo  {journal} {Phys. Rev. B}\ }\textbf {\bibinfo {volume} {108}} (\bibinfo {year} {2023}),\ 10.1103/physrevb.108.125143}\BibitemShut {NoStop}%
\bibitem [{\citenamefont {Liao}\ and\ \citenamefont {Galitski}(2018)}]{liao2018nonlinear}%
  \BibitemOpen
  \bibfield  {author} {\bibinfo {author} {\bibfnamefont {Y.}~\bibnamefont {Liao}}\ and\ \bibinfo {author} {\bibfnamefont {V.}~\bibnamefont {Galitski}},\ }\href@noop {} {\bibfield  {journal} {\bibinfo  {journal} {Phys. Rev. B}\ }\textbf {\bibinfo {volume} {98}},\ \bibinfo {pages} {205124} (\bibinfo {year} {2018})}\BibitemShut {NoStop}%
\bibitem [{\citenamefont {Swingle}\ and\ \citenamefont {Chowdhury}(2017)}]{swingle2017slow}%
  \BibitemOpen
  \bibfield  {author} {\bibinfo {author} {\bibfnamefont {B.}~\bibnamefont {Swingle}}\ and\ \bibinfo {author} {\bibfnamefont {D.}~\bibnamefont {Chowdhury}},\ }\href@noop {} {\bibfield  {journal} {\bibinfo  {journal} {Phys. Rev. B}\ }\textbf {\bibinfo {volume} {95}},\ \bibinfo {pages} {060201} (\bibinfo {year} {2017})}\BibitemShut {NoStop}%
\bibitem [{\citenamefont {He}\ and\ \citenamefont {Lu}(2017)}]{he2017characterizing}%
  \BibitemOpen
  \bibfield  {author} {\bibinfo {author} {\bibfnamefont {R.-Q.}\ \bibnamefont {He}}\ and\ \bibinfo {author} {\bibfnamefont {Z.-Y.}\ \bibnamefont {Lu}},\ }\href@noop {} {\bibfield  {journal} {\bibinfo  {journal} {Phys. Rev. B}\ }\textbf {\bibinfo {volume} {95}},\ \bibinfo {pages} {054201} (\bibinfo {year} {2017})}\BibitemShut {NoStop}%
\bibitem [{\citenamefont {Chen}(2016)}]{chen2016universal}%
  \BibitemOpen
  \bibfield  {author} {\bibinfo {author} {\bibfnamefont {Y.}~\bibnamefont {Chen}},\ }\href@noop {} {\bibfield  {journal} {\bibinfo  {journal} {arXiv preprint arXiv:1608.02765}\ } (\bibinfo {year} {2016})}\BibitemShut {NoStop}%
\bibitem [{\citenamefont {Georges}\ \emph {et~al.}(1996)\citenamefont {Georges}, \citenamefont {Kotliar}, \citenamefont {Krauth},\ and\ \citenamefont {Rozenberg}}]{DMFT_RMP}%
  \BibitemOpen
  \bibfield  {author} {\bibinfo {author} {\bibfnamefont {A.}~\bibnamefont {Georges}}, \bibinfo {author} {\bibfnamefont {G.}~\bibnamefont {Kotliar}}, \bibinfo {author} {\bibfnamefont {W.}~\bibnamefont {Krauth}}, \ and\ \bibinfo {author} {\bibfnamefont {M.~J.}\ \bibnamefont {Rozenberg}},\ }\href {\doibase 10.1103/RevModPhys.68.13} {\bibfield  {journal} {\bibinfo  {journal} {Rev. Mod. Phys.}\ }\textbf {\bibinfo {volume} {68}},\ \bibinfo {pages} {13} (\bibinfo {year} {1996})}\BibitemShut {NoStop}%
\bibitem [{\citenamefont {Bethe}(1935)}]{bethe1935statistical}%
  \BibitemOpen
  \bibfield  {author} {\bibinfo {author} {\bibfnamefont {H.~A.}\ \bibnamefont {Bethe}},\ }\href@noop {} {\bibfield  {journal} {\bibinfo  {journal} {Proceedings of the Royal Society of London. Series A-Mathematical and Physical Sciences}\ }\textbf {\bibinfo {volume} {150}},\ \bibinfo {pages} {552} (\bibinfo {year} {1935})}\BibitemShut {NoStop}%
\bibitem [{\citenamefont {Tsuji}\ \emph {et~al.}(2017)\citenamefont {Tsuji}, \citenamefont {Werner},\ and\ \citenamefont {Ueda}}]{Tsuji_2017}%
  \BibitemOpen
  \bibfield  {author} {\bibinfo {author} {\bibfnamefont {N.}~\bibnamefont {Tsuji}}, \bibinfo {author} {\bibfnamefont {P.}~\bibnamefont {Werner}}, \ and\ \bibinfo {author} {\bibfnamefont {M.}~\bibnamefont {Ueda}},\ }\href {\doibase 10.1103/physreva.95.011601} {\bibfield  {journal} {\bibinfo  {journal} {Phys. Rev. A}\ }\textbf {\bibinfo {volume} {95}} (\bibinfo {year} {2017}),\ 10.1103/physreva.95.011601}\BibitemShut {NoStop}%
\bibitem [{\citenamefont {Freericks}\ and\ \citenamefont {Zlati\ifmmode~\acute{c}\else \'{c}\fi{}}(2003)}]{Freericks_Zlatic_2003}%
  \BibitemOpen
  \bibfield  {author} {\bibinfo {author} {\bibfnamefont {J.~K.}\ \bibnamefont {Freericks}}\ and\ \bibinfo {author} {\bibfnamefont {V.}~\bibnamefont {Zlati\ifmmode~\acute{c}\else \'{c}\fi{}}},\ }\href {\doibase 10.1103/RevModPhys.75.1333} {\bibfield  {journal} {\bibinfo  {journal} {Rev. Mod. Phys.}\ }\textbf {\bibinfo {volume} {75}},\ \bibinfo {pages} {1333} (\bibinfo {year} {2003})}\BibitemShut {NoStop}%
\bibitem [{\citenamefont {Eckstein}\ and\ \citenamefont {Werner}(2011)}]{Eckstein_Werner_2011}%
  \BibitemOpen
  \bibfield  {author} {\bibinfo {author} {\bibfnamefont {M.}~\bibnamefont {Eckstein}}\ and\ \bibinfo {author} {\bibfnamefont {P.}~\bibnamefont {Werner}},\ }\href {\doibase 10.1103/PhysRevLett.107.186406} {\bibfield  {journal} {\bibinfo  {journal} {Phys. Rev. Lett.}\ }\textbf {\bibinfo {volume} {107}},\ \bibinfo {pages} {186406} (\bibinfo {year} {2011})}\BibitemShut {NoStop}%
\bibitem [{\citenamefont {Stan}\ \emph {et~al.}(2009)\citenamefont {Stan}, \citenamefont {Dahlen},\ and\ \citenamefont {Van~Leeuwen}}]{stan2009time}%
  \BibitemOpen
  \bibfield  {author} {\bibinfo {author} {\bibfnamefont {A.}~\bibnamefont {Stan}}, \bibinfo {author} {\bibfnamefont {N.~E.}\ \bibnamefont {Dahlen}}, \ and\ \bibinfo {author} {\bibfnamefont {R.}~\bibnamefont {Van~Leeuwen}},\ }\href@noop {} {\bibfield  {journal} {\bibinfo  {journal} {J. Chem. Phys.}\ }\textbf {\bibinfo {volume} {130}},\ \bibinfo {pages} {224101} (\bibinfo {year} {2009})}\BibitemShut {NoStop}%
\end{thebibliography}%

\end{document}